\documentclass[aps,prd,prabib,showpacs,nofootinbib,10pt]{revtex4}
\usepackage{graphicx} \usepackage{amsmath} \usepackage{amssymb}
\usepackage{amsfonts} \usepackage{bm}
\usepackage{array}
\usepackage{siunitx}
\usepackage[singlelinecheck=false]{caption}
\usepackage{ytableau}
\usepackage{multirow}
\usepackage{mathtools}
\usepackage{fancyhdr}

\usepackage{tikz}

\usepackage{xcolor}

\fancypagestyle{specialfooter}{%
\fancyhf{}

\fancyfoot[R]{ \noindent\fbox{%
\parbox{\textwidth}{%
{\footnotesize \bf \tiny{This version of the article has been accepted for publication, after peer review (when applicable) but is not the Version of Record and does not reflect post-acceptance
improvements, or any corrections. The Version of Record is available online at: http://dx.doi.org/10.1140/epjd/s10053-023-00758-7. Use of this Accepted Version is subject to the publisher’s Accepted
Manuscript terms of use https://www.springernature.com/gp/open-research/policies/acceptedmanuscript- terms}}
}
}}
}

\begin{document}

\newcommand{\be}{\begin{equation}} \newcommand{\ee}{\end{equation}}
\newcommand{\bea}{\begin{eqnarray}}\newcommand{\eea}{\end{eqnarray}}

\title{Quantum walk  search  by  Grover search on coin space}

\author{Pulak Ranjan Giri} \email{pu-giri@kddi-research.jp}

\affiliation{KDDI Research,  Inc.,  Fujimino-shi, Saitama, Japan}

%%%%%%%%%%%%%%%%%%%%%%%%%
\begin{abstract} 
Quantum walk followed by some  amplitude amplification  technique  has been successfully used to search for marked  vertices  on  various graphs.  Lackadaisical quantum walk can  search for target vertices  on graphs without  the help of any additional amplitude amplification  technique.  These studies either exploit   AKR or  SKW coin to distinguish  the marked vertices from the unmarked vertices.  The success of AKR  coin  based  quantum walk search algorithms  highly depend  on the arrangements of the set of marked vertices  on the graph.  For example,  it fails to find  adjacent vertices,  diagonal  vertices   and other exceptional configurations of vertices on a two-dimensional periodic  square lattice and on other graphs. These coins also suffer from low success probability  while searching for marked  vertices on a  one-dimensional periodic lattice and  on other graphs  for certain  arrangements for marked vertices.   In this article,  we  propose   a modified   coin for the   lackadaisical  quantum walk search. It allows   us to    perform  quantum walk search for  the marked vertices by  doing   Grover search on the coin space.  Our model  finds the marked vertices   by searching  the self-loops   associated with the marked  vertices.   It  can  search  for marked vertices   irrespective of their  arrangement on the graph with high success probability. For all analyzed  arrangements of the marked vertices the time complexity for  1d-lattice and  2d-lattice are    $\mathcal{O}(\frac{N}{M})$ and    $\mathcal{O}\left(\sqrt{ \frac{N}{M}\log \frac{N}{M}}\right)$  respectively with constant and high success probability.

 \end{abstract}


\date{\today}

\maketitle\thispagestyle{specialfooter}

\maketitle 

%\newpage

%\tableofcontents

\section{Introduction} \label{in}
%%%%%%%%%%%%%%%%%%%%%%%%%

Quantum computer, which  works  on the principles  of quantum mechanics,   is supposed to be  more efficient and powerful  than the most powerful classical computer  we use today.    The inefficiency of the classical  computer  to simulate  some of the  quantum systems even  with a modest size   together with the  ever decreasing size of the  transistor, leading to unavoidable  quantum effects,   are  among   the important reasons why    quantum computation  is  considered  as an alternative  form of   computing.

One of the widely studied quantum algorithms, which  shows the  superiority of the quantum computing,  is Grover's  quantum search algorithm  \cite{grover1, grover2,radha1}. In this algorithm, an unsorted database of size $N$ is given,  where some of the  elements  are marked.   The task of the algorithm  is to find the marked elements with  constant   and    high  success probability.  Classically,  one needs  to check each element  one after another to find out the marked element. In the worst case scenario, classical exhaustive search needs $\mathcal{O}(N)$  time to find out the marked element with certainty.  Grover's algorithm can do the job better by finding the marked vertices in less time. In particular, it  can search  a  marked vertex  in $\mathcal{O}(\sqrt{N})$  time, which is  quadratically faster  \cite{giri}   than  the exhaustive classical search.

Sometimes, there can be restrictions  on the way  the elements of a database can be accessed.  Graph is one  such case, where  one can impose a restriction so that  only transition from one vertex to the   nearest neighbour  vertices are  allowed at a time. Grover algorithm is not suitable for searching on a graph, because  it requires  $\mathcal{O}(\sqrt{N})$   time for the  iterations  and $\mathcal{O}(\sqrt{N})$  time to perform all the  reflections in each iteration,  reducing the quadratic speed  to the classical exhaustive search speed $\mathcal{O}(N)$ \cite{beni}. 
On the other hand  there are quantum methods, which  can search for vertices on  graphs   faster than their classical counterpart \cite{childs,amba2,meyer,amba4}.  For example, recursive  algorithm \cite{amba1}  together with the  amplitude amplification   \cite{brassard}  can  search a  marked  vertex   on   a  two-dimensional grid   in  $\mathcal{O}(\sqrt{N}\log^2 N)$ time, reaching optimal speed of   $\mathcal{O}(\sqrt{N})$  on   cubic  lattices  of  dimensions  more than two. Quantum walk(QW)  \cite{portugal} is another approach which can search for a  marked vertex on a graph faster than the classical method.   For example,  it can search for a vertex   on   a two-dimensional square lattice  in  $\mathcal{O} (\sqrt{N} \log N )$  time, while reaching optimal speed   of  $\mathcal{O} (\sqrt{N} )$     on    $d \geq 3$-dimensional  lattice   \cite{amba2,childs1}.   
On   two-dimensional  square lattice  an  improvement by a factor  of  $\mathcal{O} ( \sqrt{\log N })$    in the  time complexity is  achieved  by  using different techniques \cite{tulsi, amba3} in quantum walk searching.  On the other hand,   lackadaisical quantum walk  \cite{wong1,wong2,wong3}   can  search for a marked  vertex   in   $\mathcal{O} ( \sqrt{N \log N })$ time without the need for any additional technique.  Very recently, we numerically showed with the help of discrete-time lackadaisical quantum walk  search  that it is possible to achieve optimal speed of $\mathcal{O} (\sqrt{\frac{N}{M}})$ on two-dimensional periodic lattice, provided the grid is attached with extra long range edges given by the  Hanoi network \cite{giri3}.
Similar conclusion is drawn for searching a single marked vertex   in continuous-time quantum walk  by using completely different long range edges on a two-dimensional grid \cite{tomo}.

Despite the huge success of the  Grover oracle based quantum walk  as well as the  lackadaisical quantum walk  search  for  finding single marked vertex   on various graphs,  generalisation of these  search algorithms  to  multiple marked  vertices   is severely restricted. 
Multiple vertices of a two-dimensional periodic lattice can  be searched  only for certain arrangements   of the marked vertices \cite{rivosh,saha, nahi}.  
Similarly,  lackadaisical quantum walk  can   search one of the  $M$ vertices  in   $\mathcal{O} (\sqrt{\frac{N}{M} \log \frac{N}{M}})$ time \cite{giri2} with  constant  success probability for  the set of marked vertices, which are not adjacent to each other.  However, two adjacent vertices can not be found  by these discrete-time quantum walk search algorithms.  Marked vertices arranged  along the diagonal of a two-dimensional grid also  can not be found.  Similarly, set of marked vertices arranged as a block of  $2k \times l$ or  $k \times 2l$, for any positive $k, l$, can not found  by discrete-time quantum walk.  Only  for the set of vertices  arranged in a block of size $k \times l$, for $k,l$ both being odd, quantum walk can be used to  search any  of the marked vertices in the block.

In this article  we  propose a modification to the lackadaisical quantum walk search, which can  find     multiple vertices arranged in a cluster of any form on a two-dimensional periodic square lattice.  Instead of flipping the sign of phases  of  all the basis  states of the coin space associated with the marked vertices we only flip the sign of the self-loops  associated with the marked vertices.   Quantum walk search is then Grover search on the coin space followed by flip-flop shift operation.  Our model can search for marked vertices   arranged in a cluster of  any size and shape.

%----------------------------------------------------------------------------------------------------
\begin{figure}[h!]
  \centering
     \includegraphics[width=0.80\textwidth]{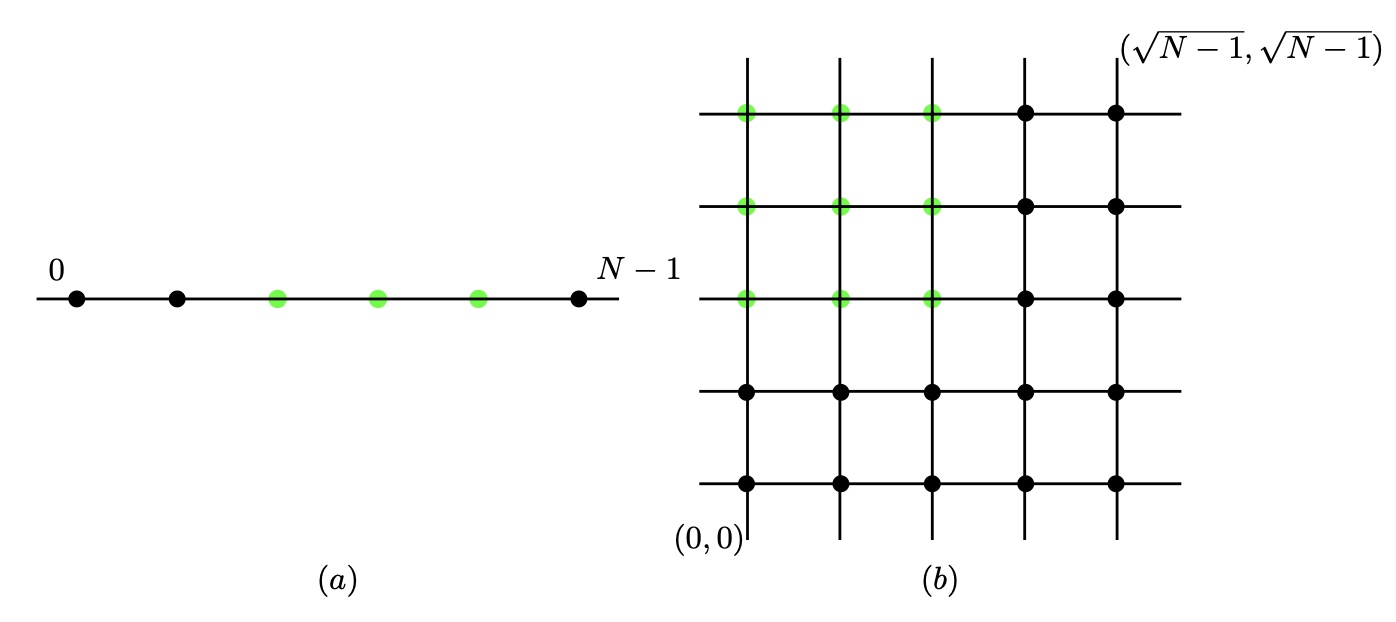}
          
       \caption{ (Color online) (a)  One-dimensional grid of  size   $N$   with periodic boundary conditions, and  (b)  two-dimensional grid of  size   $\sqrt{N} \times \sqrt{N}$   with periodic boundary conditions with a green coloured cluster of marked  vertices.}
\end{figure}
%----------------------------------------------------------------------------------------------------

We   arrange  this article  in the following fashion:  A  discussion  on the  quantum walk  search  with  the  proposed modification of oracle  is provided in section    \ref{qw}.   In section \ref{1D} and  \ref{2D}  we discuss  lackadaisical quantum walk search,  equipped with the proposed oracle,    on one and two-dimensional   periodic  lattice  respectively     and  finally in section \ref{con} we conclude.

%%%%%%%%%%%%%%%%%%%%%%%%%%%%%%%%%%%%%%%%%%%%%%%%%%%%%%%%
\section{ Quantum walk   search by Grover search} \label{qw}
%%%%%%%%%%%%%%%%%%%%%%%%%
In this section we discuss  our proposed modification to the lackadaisical   quantum walk  search, which can be successfully exploited to find  multiple marked vertices irrespective of  their  arrangements on the graph.  We consider a    vertex transitive graph $G(V, E)$  of degree $d$  with  $|V|$ vertices and $|E|$ edges.
In  discrete-time  version of the quantum walk,  the  walker moves  from one vertex of  the  graph to the  nearest  neighbour vertices  depending on  the state of the walker(coin).  
The state of the graph  is represented as an element of  the Hilbert space   $\mathcal{H}_G =  \mathcal{H}_C \times \mathcal{H}_V$, where   $\mathcal{H}_V$ is the space of vertices and 
$\mathcal{H}_C$ is the  space of coin states.  If there are $|V|= N$ vertices  in the graph,  then the dimension of   $\mathcal{H}_V$ is  $N$.  There is  one self-loop at each vertex of the graph, which allows some fraction of the amplitude to stay at the same  vertex even after the shift operation. So the dimension of  the coin space  $\mathcal{H}_C$ is $d+1$.  For the quantum walk search we need to prepare an initial state  of the graph, which we choose  as   an  equal superposition of basis states  on both coin and vertex space respectively
\begin{eqnarray}
 |\psi_{in}\rangle = |\psi_{c}\rangle \otimes  |\psi_{v}\rangle =  \frac{1}{\sqrt{d +a}}  \left(\sum^{d-1}_{i =0}  |x_c^{i}\rangle + \sqrt{a}|\mbox{loop}\rangle \right) \otimes \frac{1}{\sqrt{N}} \sum^{N-1}_{i=0} |x_v^i \rangle\,.
 \label{in}
\end{eqnarray}
The  initial state   $|\psi_{in}\rangle$   is then  evolved by  a  unitary operator $\mathcal{U}$   multiple times so that the marked state is obtained with hight success probability.      $\mathcal{U}$ is composed of  the coin  operator  followed by the   shift operator   respectively.   Standard shift operator moves the walker to the nearest  neighbour vertices keeping the direction of movement same.  For the quantum walk search it fails to increase the success probability of the marked vertex.  There is another  shift operator $S$, known as the  flip-flop shift operator, which moves the walker to the nearest neighbour  vertices and then  turn around its direction.

One of the important things  in  quantum search  is to   distinguish   the marked  vertices   from the unmarked vertices.  It is usually done  by  modifying  the   coin operator  $\mathcal{C}$    as 
\begin{eqnarray}
\mathcal{C}=  C_+ \otimes \left( \mathbb{I}_{N \times N} -    \sum_{i=1}^M |t_i \rangle \langle t_i |  \right) + C_- \otimes  \sum_{i=1}^M |t_i \rangle \langle t_i | \,, 
\label{qc}
\end{eqnarray}
where    $C_+$ and $C_-$  are  the coin operators   for the unmarked and marked vertices respectively and    $|t_i \rangle$s  are  $M$   marked vertex states.   One of the widely used  modified coins   \cite{amba3}
\begin{eqnarray}
\mathcal{C}_{AKR}=  C_0 \otimes \left( \mathbb{I}_{N \times N} -   2 \sum_{i=1}^M |t_i \rangle \langle t_i |  \right) \,, 
\label{qc1}
\end{eqnarray}
can be obtained by replacing $C_+ =  -C_ - =  C_0$  in eq. (\ref{qc}).  Another type of modified coin  \cite{she}
\begin{eqnarray}
\mathcal{C}_{SKW}=  C_0 \otimes \left( \mathbb{I}_{N \times N} -    \sum_{i=1}^M |t_i \rangle \langle t_i |  \right) - \mathbb{I}_{d+1 \times d+1} \otimes  \sum_{i=1}^M |t_i \rangle \langle t_i | \,, 
\label{qc2}
\end{eqnarray}
can be obtained by replacing $C_+ =  C_0$ and   $C_ - =  - \mathbb{I}$  in eq. (\ref{qc}). One of the choices  for the coin flip operator $C_0$ is the Grover diffusion operator
\begin{eqnarray}
C_0 =   \left( 2  | \psi_c \rangle \langle  \psi_c | -  \mathbb{I}_{d+1 \times d+1} \right) \,, 
\label{qop}
\end{eqnarray}
which is also used in our numerical analysis.  $\mathcal{C}_{AKR}$   coin has been frequently  used in  the discrete-time based quantum walk search algorithms to  distinguish  the marked vertices from the unmarked vertices. Like the  Grover algorithm,   $\mathcal{C}_{AKR}$  coin  also inverts the sign of the marked vertices followed by the  Grover diffusion operator in the coin space.  It  is  very much  useful to search a single marked vertex and some specifically arranged multiple marked  vertices  on a graph. However,  it  fails to  find  multiple  marked vertices which form  exceptional configurations \cite{men}.

In this article we propose a modification of  the  coin  in eq. (\ref{qc1})  as 
\begin{eqnarray}
\mathcal{C}_{G}=  \left(C_0 \otimes \mathbb{I}_{N \times N} \right)  \left( \mathbb{I}_{d+1 \times d+1}  \otimes \mathbb{I}_{N \times N}-   2 \sum_{i=1}^M | \mbox{loop}, t_i \rangle \langle  \mbox{loop}, t_i |  \right) \,, 
\label{qc3}
\end{eqnarray}
where we  only flip the sign of those   $| \mbox{loop} \rangle$ states    which are  attached with  the marked vertices.  The key difference between   $\mathcal{C}_{AKR}$   and the proposed coin  is that    $\mathcal{C}_{AKR}$  flips the sign of all the $d+1$ basis states of the coin space attached to the marked vertices, while our coin $\mathcal{C}_{G}$ only flip the sign of the self-loop states  attached with the marked vertices. 
The evolution operator  for our  quantum walk search is of the form
\begin{eqnarray}
 \mathcal{U} = S \mathcal{C}_{G}\,,
\label{uqw}
\end{eqnarray}
where the flip-flop shift operator   $S$  acts on both the vertex space and the coin space respectively. 
%----------------------------------------------------------------------------------------------------
\begin{figure}[h!]
  \centering
     \includegraphics[width=0.80\textwidth]{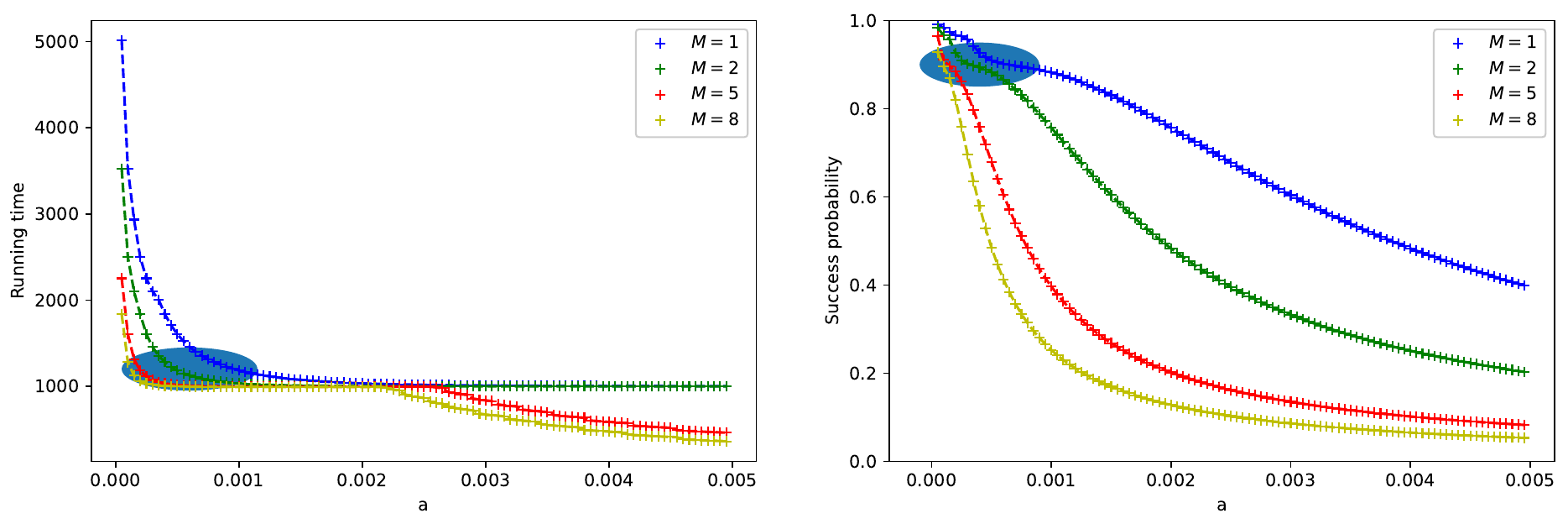}
          
       \caption{(Color online) (a) Variation of the  running time for the first peak of success   probability  for  $M =1$(blue), $2$(green), $5$(red) and $8$(yellow) adjacent targets and (b) corresponding success probability   as a function of the self-loop weight
       $Na$  for    $N= 1000$  one-dimensional periodic lattice.  
       }

\end{figure}
%----------------------------------------------------------------------------------------------------

 %----------------------------------------------------------------------------------------------------
\begin{figure}[h!]
  \centering
     \includegraphics[width=0.80\textwidth]{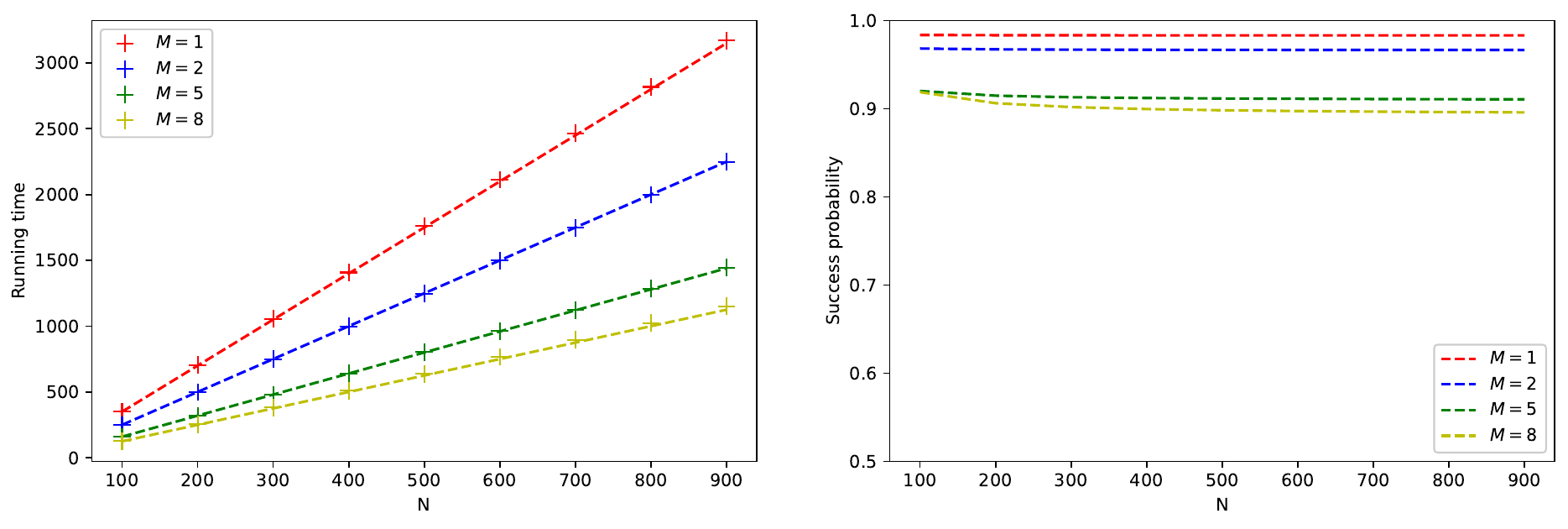}
          
       \caption{(Color online) (a) Running time of  quantum walk search for $M =1$(red), $2$(blue), $5$(green)  and $8$(yellow)  targets on a one-dimensional periodic grid with  self-loop weight $a= 0.1/N $  and (b)  the  corresponding success probability    as a function of  number of vertices   $N$. }

\end{figure}
%----------------------------------------------------------------------------------------------------

This   shift operator  moves the  quantum walker   from a given vertex to nearest neighbour vertices   depending  on  the state of the  quantum coin  and  flips   the state of the coin.   
After   repeated application of  $\mathcal{U}$ the final  state  $|\psi_{f}\rangle =    \mathcal{U} ^{t} |\psi_{in }\rangle$   has     $\mathcal{O}(1)$   success probability of being in the marked vertices.

%%%%%%%%%%%%%%%%%%%%%%%%%%%%%%%%%%%%%%%%%%%%%%%%%%%%%%%%
\section{One-dimensional periodic lattice} \label{1D}
%%%%%%%%%%%%%%%%%%%%%%%%%
In this section,  we consider  a  one-dimensional  lattice with periodic boundary conditions   for  the quantum walk search.   In fig. 1(a) an example of a one-dimensional periodic lattice is  depicted for  $N = 6$  vertices.  It is one of the simplest graph to study spacial search algorithm.  Although, the time complexity  $\mathcal{O}(N)$  \cite{giri2} for the  spatial search on a  one-dimensional periodic lattice does not improve compared to the exhaustive classical search, it  is  quadratically faster than the corresponding classical random walk search.  Lackadaisical quantum walk  with Grover oracle can search a marked  vertex  with  success probability of about   $0.75$.  We have  observed that using our  coin in eq.  (\ref{qc3}) we can boost the success probability  to $\approx 1.0$  with the same time complexity  $\mathcal{O}(N)$  but increased   pre-factor.

The    initial state of the quantum coin  at each vertex of the lattice  is  given by:
\begin{eqnarray}
|\psi_c \rangle   =  \frac{1}{\sqrt{2+ a}} \left( |c_x^{+}\rangle + |c_x^{-}\rangle+ 
 \sqrt{a}| \mbox{loop}\rangle \right) \,,
\label{1Dcstate}
\end{eqnarray}
and the initial state of  the  vertex space of the graph  is given by 
\begin{eqnarray}
|\psi_{v}\rangle =   \frac{1}{\sqrt{N}} 
\sum^{N}_{{v_x} = 1} |v_x\rangle \,.
\label{1Din}
\end{eqnarray}
We use the Grover diffusion operator  in our analysis  
\begin{eqnarray}
C_0 = 2 |\psi_c \rangle \langle \psi_c | - \mathbb{I}_3\,. 
\label{2Dcgrov}
\end{eqnarray}
The  flip-flop shift operator  $S$   is  given by
\begin{eqnarray} 
S  =  \sum^{ N}_{v_x=1}   \bigg[ |c_x^{-} \rangle \langle  c_x^{+} | \otimes | v_x +1 \rangle \langle  v_x  | 
+   |c_x^{+} \rangle \langle  c_x^{-} | \otimes | v_x -1  \rangle \langle  v_x  |   +   |\mbox{loop} \rangle \langle  \mbox{loop} | \otimes | v_x  \rangle \langle  v_x  |    \bigg]\,.
\label{fshift1}
\end{eqnarray}
Let us now discuss how the   evolution operator $\mathcal{U}$ acts on  a general state of the  graph.  Consider a general state of the coin space  $|\psi_1 \rangle   =    \left( b_1|c_x^{+}\rangle + b_2|c_x^{-}\rangle+  b_3| \mbox{loop}\rangle \right)$.    Firstly  the  coin operator  $\mathcal{C}_{G}$  acts as:
 \begin{eqnarray} \label{1dc1}
\mathcal{C}_{G}|\psi_1 \rangle  | v_x \rangle &=&   \left(2 |\psi_c \rangle \langle \psi_c | - \mathbb{I}_3 \right)  \left( \mathbb{I}_3- 2 |\mbox{loop} \rangle \langle \mbox{loop} | \right) |\psi_1 \rangle  | v_x \rangle\,,  \mbox{for} ~~ v_x = t_i\\ 
&=&   \left(2 |\psi_c \rangle \langle \psi_c | - \mathbb{I}_3 \right)   |\psi_1 \rangle  | v_x \rangle\,, ~~~~~~~~~~~~~~~~~~~~~~~~~ \mbox{for} ~~ v_x \neq t_i
\label{1dc2}
\end{eqnarray}
Then the flip-flop shift operator acts as: 
\begin{eqnarray}
S| c_x^{+} \rangle| v_x \rangle &=& | c_x^{-} \rangle| v_x + 1\rangle\,, \\
S| c_x^{-} \rangle| v_x \rangle &=& | c_x^{+} \rangle| v_x - 1\rangle\,, \\
S| \mbox{loop} \rangle| v_x \rangle &=& | \mbox{loop} \rangle| v_x \rangle\,.
\label{2Dcgrov}
\end{eqnarray}
Similar to the partial Grover search algorithm  \cite{giri} we can think of coin spaces which are attached to the marked vertices as target blocks and rest of  the coin spaces as non-target blocks. 
Then, at the  marked vertices   $\mathcal{C}_{G}$ acts as the Grover iterator on  the coin space(target blocks)  with the  self-loop as the marked element  as can be seen from eq. (\ref{1dc1}) and at the other vertices it acts  on the coin space(non-target blocks) as if there is no marked element as  can be seen  from eq. (\ref{1dc2}). 

{\it Experimental results:} Performance of the lackadaisical quantum walk search depends on the value of the self-loop weight  $a$.  A suitable choice for the self-loop weight is necessary for obtaining  high success probability and a running time with reasonably small pre-factor.  A simulation for a one-dimensional periodic lattice with $N=1000$ has been  performed.  Variation of the running time and success probability of the first-peak has been provided in fig. 2 (a) and (b) respectively.  We see that the running time in fig. 2 (a) sharply increases 
as the self-loop weight $a$ goes towards zero. However the region inside the blue area the running time is reasonably small and  the corresponding  success probability as shown by blue region in fig. 2 (b)  is  very high.  Of course the success probability can be increased close to $1$ at the expenses of the running time constant pre-factor.  In our analysis we have chosen the self-loop weight in the blue patch region for further analysis. In fig. 3 (a) and (b)  running time and success probability as a function of the number of vertices $N$ have  been plotted for multiple marked vertices in a cluster. Self-loop weight is fixed at  $a= 0.1/N$.  As we can see  $M =1, 2, 5, 8$ marked vertices can be searched in    $\mathcal{O}(\frac{N}{M})$ time with over $0.9$ success probability.  Note that using Grover oracle based lackadaisical quantum walk search one marked vertex  can be searched with a success probability of  not more than   $0.75$ \cite{giri2}, however using our modified oracle success probability is  $\sim 0.98$ as depicted by red  curve in fig. 3 (b).  

%%%%%%%%%%%%%%%%%%%%%%%%%%%%%%%%%%%%%%%%%%%%%%%%%%%%%%%%
\section{Two-dimensional periodic lattice} \label{2D}
%%%%%%%%%%%%%%%%%%%%%%%%%
In this section we discuss  how we can search for target vertices  on a  $2$-dimensional periodic  lattice   by doing Grover search  on coin space to find the self-loop.  In fig. 1(b)  a   periodic $2$-dimensional   lattice    of size   $\sqrt{N} \times \sqrt{N}$ is displayed, where    $N$    elements of  a database are represented  as  the   $N$   vertices     $(x, y)$   of the graph.   Each vertex is connected to  five   edges, four of them come from the four edges of the $2$-dimensional lattice and remaining  one comes from the self-loop of lackadaisical quantum walk.   

%----------------------------------------------------------------------------------------------------
\begin{figure}[h!]
  \centering
     \includegraphics[width=0.80\textwidth]{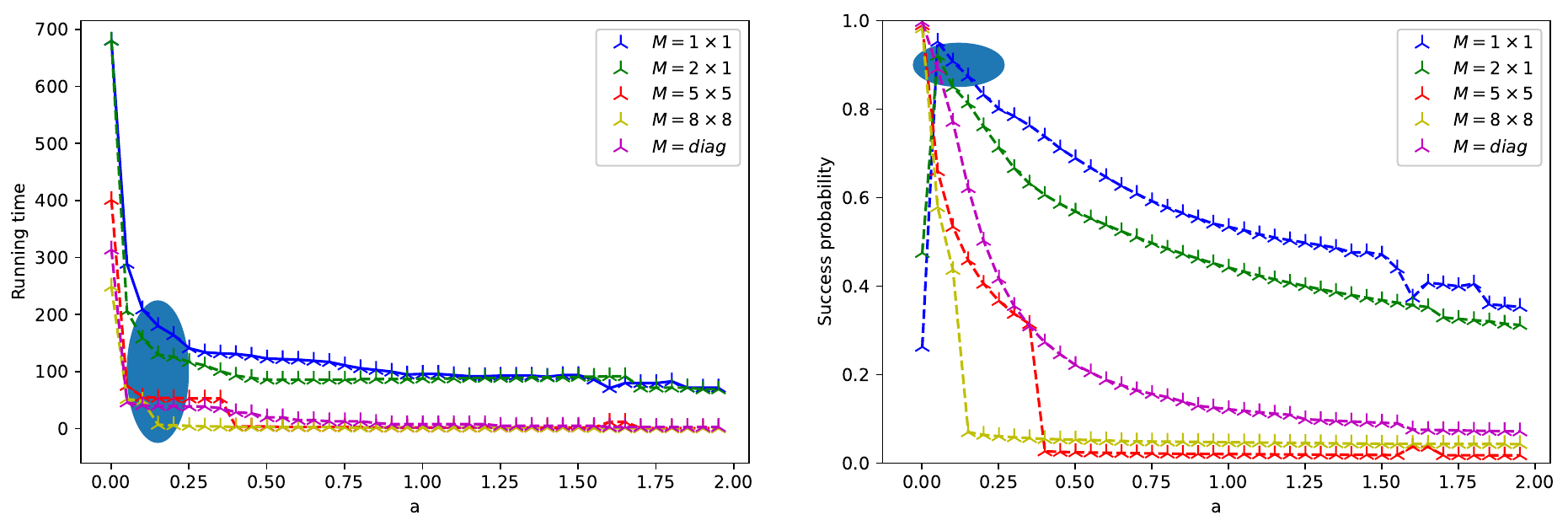}
          
      \caption{(Color online) (a) Variation of the  running time for the first peak of  probability  for a set of adjacent targets of sizes    $M =1 \times 1$(blue), $2 \times 1$(green), $5 \times 5$(red), $8 \times 8$(yellow) and $M= \mbox{diag}$(purple)   and (b) corresponding success probability   as a function of the self-loop weight
       $a$  for     $ \sqrt{N} \times  \sqrt{N}= 40 \times 40$  two-dimensional periodic lattice.  
       }

\end{figure}
%----------------------------------------------------------------------------------------------------

 %----------------------------------------------------------------------------------------------------
\begin{figure}[h!]
  \centering
     \includegraphics[width=0.80\textwidth]{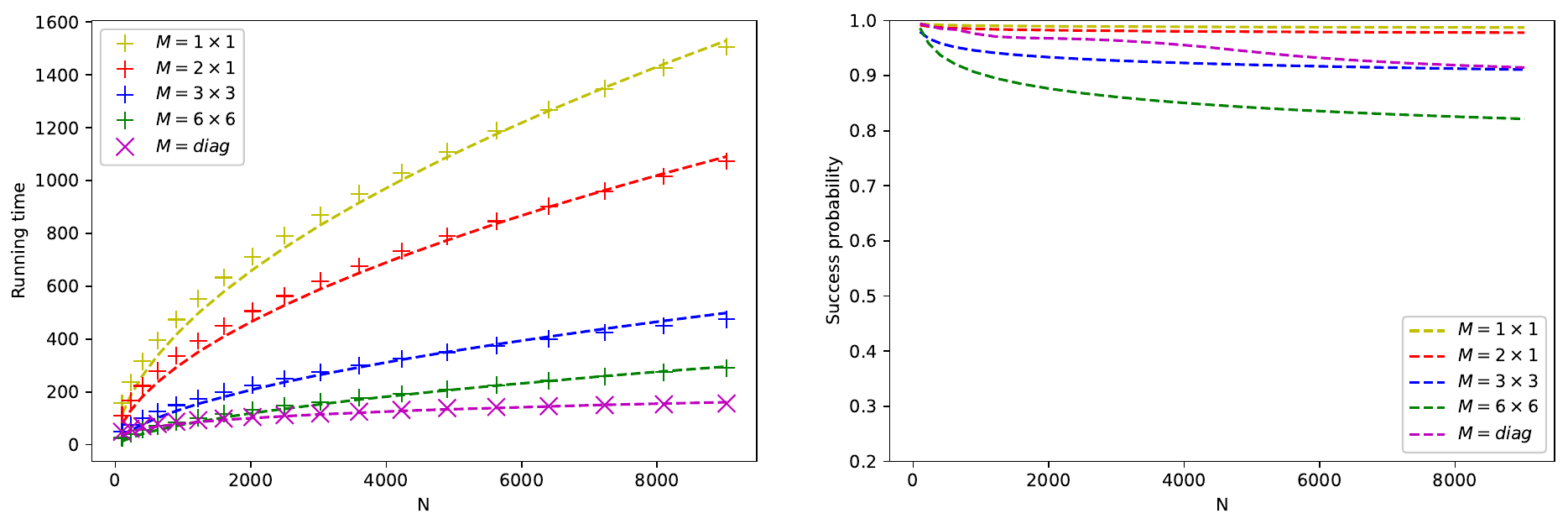}
          
       \caption{(Color online) (a) Running time of  quantum walk search for $M =1 \times 1$(yellow), $2 \times 1$(red), $3 \times 3$(blue),  $6 \times 6$(green)  and $M = diag(\mbox{purple})$  marked vertices  on a two dimensional grid with self-loop weight $a=0.01$ and (b)  the  corresponding success probability    as a function of  number of vertices  $N$.}

\end{figure}
%----------------------------------------------------------------------------------------------------

Therefore,   initial state of the quantum coin  is given by:
\begin{eqnarray}
|\psi_c \rangle   =  \frac{1}{\sqrt{4+ a}} \left( |c_x^{+}\rangle + |c_x^{-}\rangle+ 
 |c_y^{+}\rangle + |c_y^{-}\rangle   + \sqrt{a}| \mbox{loop}\rangle \right) \,.
\label{2Dcstate}
\end{eqnarray}
The initial state of  the  vertex space of the graph  is given by 
\begin{eqnarray}
|\psi_{v}\rangle =   \frac{1}{\sqrt{N}} 
\sum^{\sqrt{N}-1}_{{v_x, v_y} = 0} |v_x,  v_y\rangle \,.
\label{2Din}
\end{eqnarray}
In our quantum search problem   we choose the Grover diffusion operator  
\begin{eqnarray}
C_0 = 2 |\psi_c \rangle \langle \psi_c | - \mathbb{I}_5\,, 
\label{2Dcgrov}
\end{eqnarray}
for  the rotation of the  quantum coin state.    The shift operator  $S$   associated with  the four standard edges and a self-loop  of the $2$-dimensional square lattice  is  given by
\begin{eqnarray} \nonumber
S =  \sum^{ \sqrt{N}-1}_{v_x, v_y=0}  && \bigg[ |c_x^{-} \rangle \langle  c_x^{+} | \otimes | v_x +1; v_y  \rangle \langle  v_x; v_y  | 
+   |c_x^{+} \rangle \langle  c_x^{-} | \otimes | v_x -1; v_y  \rangle \langle  v_x; v_y  |    \\ \nonumber
 &+&  |c_y^{-} \rangle \langle  c_y^{+} | \otimes | v_x; v_y  +1 \rangle \langle  v_x; v_y  |  
+   |c_y^{+} \rangle \langle  c_y^{-} | \otimes | v_x; v_y -1 \rangle \langle  v_x; v_y  |   \\ 
&+& | \mbox{loop} \rangle \langle  \mbox{loop} | \otimes | v_x; v_y  \rangle \langle  v_x; v_y | \bigg]\,.
\label{fshift1}
\end{eqnarray}
Similar to the one-dimensional case,    the evolution operator $\mathcal{U}$ acts on  a general state of the two-dimensional  graph.  Consider a general state of the five dimensional coin space  $|\psi_2 \rangle   =    \left( b_1|c_x^{+}\rangle + b_2|c_x^{-}\rangle+ b_3|c_y^{+}\rangle + b_4|c_y^{-}\rangle + b_5| \mbox{loop}\rangle \right)$.    Now  the  coin operator  $\mathcal{C}_{G}$  acts as:
 \begin{eqnarray} \label{2dc1}
\mathcal{C}_{G}|\psi_2 \rangle  | v_x; v_y \rangle &=&   \left(2 |\psi_c \rangle \langle \psi_c | - \mathbb{I}_3 \right)  \left( \mathbb{I}_3- 2 |\mbox{loop} \rangle \langle \mbox{loop} | \right) |\psi_2 \rangle  | v_x; v_y \rangle\,,  \mbox{for} ~~ (v_x; v_y) = t_i\\ 
&=&   \left(2 |\psi_c \rangle \langle \psi_c | - \mathbb{I}_3 \right)   |\psi_2 \rangle  | v_x; v_y \rangle\,, ~~~~~~~~~~~~~~~~~~~~~~~~~ \mbox{for} ~~ (v_x; v_y) \neq t_i
\label{2dc2}
\end{eqnarray}
and  the flip-flop shift operator acts as: 
\begin{eqnarray}
S| c_x^{+} \rangle| v_x; v_y \rangle &=& | c_x^{-} \rangle| v_x + 1; v_y\rangle\,, \\
S| c_x^{-} \rangle| v_x; v_y \rangle &=& | c_x^{+} \rangle| v_x - 1; v_y\rangle\,, \\
S| c_y^{+} \rangle| v_x; v_y \rangle &=& | c_y^{-} \rangle| v_x; v_y + 1\rangle\,, \\
S| c_y^{-} \rangle| v_x; v_y \rangle &=& | c_y^{+} \rangle| v_x; v_y - 1\rangle\,, \\
S| \mbox{loop} \rangle| v_x; v_y \rangle &=& | \mbox{loop} \rangle| v_x; v_y \rangle\,.
\label{2Dcgrov}
\end{eqnarray}
Note that our modified coin can only be implemented in lackadaisical quantum walk, because  a self-loop is necessary in our model to make Grover search in coin space.

{\it Experimental results:}  As we discussed in the introduction, there is a limitation for searching  by Grover oracle based quantum walk. Marked vertices arranged  in a form of  $2k \times l$ or 
$k \times 2l$ cluster, for  positive $k, l$,  can  not be  found  by   $\mathcal{C}_{AKR}$  based  quantum walk search algorithm.  Marked vertices which form the diagonal of a square lattice also can not be found. 
However,  our numerical results show that our modified coin  $\mathcal{C}_{G}$    can search these cluster of vertices along with any  other  arrangements  of vertices.   
 
 Since a suitable  choice of the self-loop weight is important for our search algorithm, in fig. 4 (a) and (b) we plotted the behaviour of the running time and  success probability respectively as a function of the self-loop weight for a quantum walk search on  a square lattice of size  $40 \times 40$.  Cluster of marked vertices of the form $M = 1 \times 1,
 2 \times 1, 5 \times 5$,  $8 \times 8$ and $M=$ diag(vertices along the diagonal)   are considered for the searching.   We choose  the value of the self-loop weight  in the blue region in fig. 4 (a), which corresponds to the success probability  depicted by  the  blue region in fig. 4 (b).  In  fig. 5 (a)  running time  for  finding  marked vertices arranged in  the form of  cluster  for $M = 1 \times 1, 2 \times 1,
 3 \times 3, 6 \times 6$ and $M=$ diag(vertices along the diagonal)   have  been depicted for lattice size up to $100 \times 100$ with self-loop weight $a= 0.01$.  Note  that the self-loop weight is not optimised to account for  the change in  the size of the database and the number of marked vertices, which leads to small drop in the success probability.   However, we still  have   success probability above $0.8$ in  fig. 5 (b). One can choose the self-loop weight  in such a way that the success probability is $\sim 1$, which will result in the increase in the constant factor of time complexity.  
 
 The   success probability corresponding to  the result in fig. 5(a)  is  constant as can be seen from fig. 5 (b).  Note that, the  cluster of marked vertices  $2 \times 1, 6 \times 6$ and $8 \times 8$  are exceptional configurations, which  the Grover oracle based quantum walk search algorithms is unable to find, but our proposed coin based algorithm  can efficiently find   in $\mathcal{O}\left(\sqrt{ \frac{N}{M}\log \frac{N}{M}}\right)$  time.  

%%%%%%%%%%%%%%%%%%%%%%%%%
\section{Conclusions} \label{con}
%%%%%%%%%%%%%%%%%%%%%%%%%
Quantum walk  has been very much useful  to search on  various graphs for marked vertices.  In  discrete-time quantum walk, we need to  distinguish marked vertices from the unmarked vertices.  Usually it is done by modifying  the coin operator  in such a way that it applies one  coin operator to the coin state associated with unmarked vertices and applies a different coin operator to the coin state associated with the marked vertices.  Two widely used coin operators can be found in the literature:  AKR coin, which applies  $C_0$ to the state associated with unmarked vertices and applies $-C_0$ to the state  associated with the marked vertices, and  CKW coin,  which applies  $C_0$ to the state associated with unmarked vertices and applies $- \mathbb{I}$ to the state  associated with the marked vertices.  In principle, $C_0$ can be any unitary  operator of appropriate dimensions, however Grover operator has been widely  used in the literature.
These two coins can either be used  with the standard quantum walk to search  or they can be used with the lackadaisical quantum walk to search for marked vertices.    In this article we are only considering lackadaisical quantum walk, because it does not require any additional technique to enhance  the success probability. 

Existing coins are good for single vertex searching. However, there are certain limitations while searching for multiple vertices. For example, multiple marked vertices, which are arranged in a specific form of cluster such as   $2k \times l$ or  $k \times 2l$ cluster, for  positive $k, l$,  can  not be  found  by  quantum walk with AKR coin.  Marked vertices arranged along the diagonal of a square lattice  also can not be found  by AKR coin. There  is a  wide variety of more general  exceptional configurations \cite{men} which can not be found.   These exceptional configuration affect the search by AKR coin but not by the SKW coin \cite{wong4}.  However, these two coins are equivalent to each other  for searching a single vertex on  a arbitrary-dimensional periodic  lattice  and  on a complete graph without self-loop \cite{wong1,wong4}.  Sometimes, although   we can search marked vertices, but the  success probability  is very low. For example, we observed that  success probability is limited by  $0.75$  while  searching  a single vertex on a one-dimensional  periodic lattice by lackadaisical quantum walk with AKR coin. SKW coin is  not useful to search  on a one-dimensional periodic lattice.

To overcome these limitations, we  proposed  a  coin operator  $\mathcal{C}_{G}$,  which applies Grover coin  $C_0$ to the state associated with unmarked vertices and applies $-C_0$ only to the self-loop state  associated with the marked vertices.  Numerical analysis  performed on one and two-dimensional periodic lattice shows  that the coin  $\mathcal{C}_{G}$ can find  exceptional configurations as well as any other  arrangements of the marked vertices efficiently and with very  high success probability.  It  would be interesting to exploit this algorithm to search for marked vertices on various other graphs. 

\vspace{1cm}

Data availability Statement:  The datasets generated during and/or analysed during the current study are available from the corresponding author on reasonable request.
\vspace{0.5cm}

Conflict of interest: The authors have no competing interests to declare that are relevant to the content of this article.

%%%%%%%%%%%%%%%%%%%%%%%%%%%

\end{document}